\documentstyle[11pt,newpasp,twoside,epsf]{article}

\begin{document}

\title{Coarse Estimation of Physical Parameters of Eclipsing Binaries by Means of Automative Scripting}
\author{Andrej Pr\v sa}
\affil{Faculty of Mathematics and Physics, Dept.~of Physics, University of Ljubljana, Slovenia}

\begin{abstract}
Because of GAIA's estimated harvest of $\sim 10^5$ eclipsing binaries (Munari et al.~2001) automative procedures for extracting physical parameters from observations must be introduced. We present preliminary results of automative scripting applied to 5 eclipsing binaries, for which photometric and radial velocity observations were taken from literature. Although the results are encouraging, extensive testing on a wider sample has to be performed.
\end{abstract}

\vspace{-6mm}
\section{Introduction}

Eclipsing binaries have solid grounds in today's astrophysics; due to relatively simple geometry modeling allows astronomers to extract physical parameters such as masses and radii from observed light curves and radial velocity curves in absolute units, which is very difficult (if at all possible) for other kinds of celestial objects. Modeling approaches are developing rapidly with increasing computing power of PCs. Most widely used modeling algorithms are {\tt WD} (Wilson \& Devinney 1971), {\tt WINK} (Wood 1971), {\tt FOTEL} (Hadrava 1986) and others. However, what most of these models lack is the level of automation required for scanning missions. GAIA will retrieve data for $\sim 10^5$ eclipsing binaries to $15^{\mathrm m}$ (ESA-SCI(2000)4) and fully automative approaches are absolutely necessary.

\vspace{-2mm}
\section{Automative approach}

The fact that most of the existing modeling software lacks automation isn't a simple oversight, but is a consequence of few inevitable problems one faces while seeking a solution (Wilson 1998): 1. Extraction of physical parameters of eclipsing binaries is a highly non-linear inverse problem and is as such subjected to modeling degeneracies. 2. Solution convergence may be slow or even fail because of shallow, wide minima in parameter space, numerical algorithms' inefficiency and non-linearity along with strong parameters' correlations. 3. The choice of parameters to be adjusted changes from iteration to iteration based on common sense and it is difficult to assess in advance how a certain change in parameter values changes the credibility of the entire solution.

Bearing this in mind we use Wilson--Devinney model for basis and propose an automative approach, which is subject to the following main principles: (a) The underlaying formalism of solution-seeking shouldn't be changed in any way. (b) There is more to astrophysics than parameter estimation, so any automative processing should be limited only to obtain coarse solutions. (c) The successfulness should be tested on wide and diverse types of eclipsing binaries. (d) The time cost of modeling analysis should be considerably reduced.

We refer to this process as \emph {automative scripting}. Based on photometric and spectroscopic observations, with the assumption that the orbital period $P$, the epoch HJD${}_0$ and the type of an eclipsing binary (detached, semi-detached, contact or over-contact) are known, we follow the predefined pattern of what modeling actions to perform in a particular order. Such a pattern is a \emph {script}. Our goal is to create such a script (or a set of scripts), that would lead us to a coarse modeling solution without any interactive approach.

\vspace{-3mm}
\section {Preliminary test results}

We have performed some preliminary testing of automative scripting on dozen different eclipsing binaries taken from literature. Fig. 1 shows the results of automative scripting performed on 5 stars given in Table 1. The measurements consist of 100--1600 data points per filter in photometry and 10--100 data points in radial velocities. A typical time scale for the complete process of automative scripting on a 1.4GHz processor is $\sim$35 seconds. Table 2 contains basic extracted parameters, which may be compared against adopted values given in Table 3.

\vspace{-4mm}
\begin{table}
\begin{center}
\begin{tabular}{cccl}
\hline
\rule[-0.9ex]{0pt}{3ex}
Star Name & Type         & Available data                & References               \\
\hline
\rule[-0.9ex]{0pt}{3ex}
BH Vir    & Detached     & $B$, $V$, $u$, $b$, $v$, $y$, & Clement et al.~(1997)    \\
          &              & both RVs                      &                          \\
GK Dra    & Detached,    & $B$, $V$, both RVs            & Zwitter (2002) \\
& $\delta$-Sct 2$^{\mathrm{ndary}}$ &                   &                          \\
TY Boo    & Over-contact & $B$, $V$, both RVs            & Milone et al.~(1991)     \\
UV Leo    & Detached     & $B$, $V$, both RVs            & Frederic \& Etzel (1996) \\
\rule[-0.9ex]{0pt}{3ex}
V505 Per  & Detached   & $b_T$, $H_p$, $v_T$, both RVs & Munari et al.~(2001)      \\
\hline
\end{tabular}
\end{center}
\vspace{-3mm}
\caption{A list of chosen test stars that were used to assess the automative scripting successfulness. Notation remarks: $U$, $B$ and $V$ stand for Johnson, $u$, $b$, $v$, $y$ for Str\"omgren, $b_T$, $v_T$ and $H_p$ for Tycho/Hipparcos and $I_C$ for Cousins' filters. RVs are radial velocities.}
\end{table}

Few remarks should be made: 1. The solutions were reached without any human intervention. 2. The solution degeneracy is reduced by using the method of multiple subsets (Wilson 1998) instead of exclusive use of differential corrections. 3. Further degeneracy elimination is possible because of the presence of radial velocities, which along with photometric data uniquely determine the semi-major axis and both components' masses. Because of GAIA's RVS capabilities we expect tremendous improvements to the field of eclipsing binaries.

\vspace{-3mm}
\section {Discussion}

Note the discrepancy between calculated and adopted potentials $\Omega_1$ and $\Omega_2$ in Tables 2 and 3: this is a consequence of the degeneracy between the inclination and stellar radii ($\propto \Omega^{-1}$), the direct example of how blindly trusting\linebreak

\begin{figure}
\plottwo{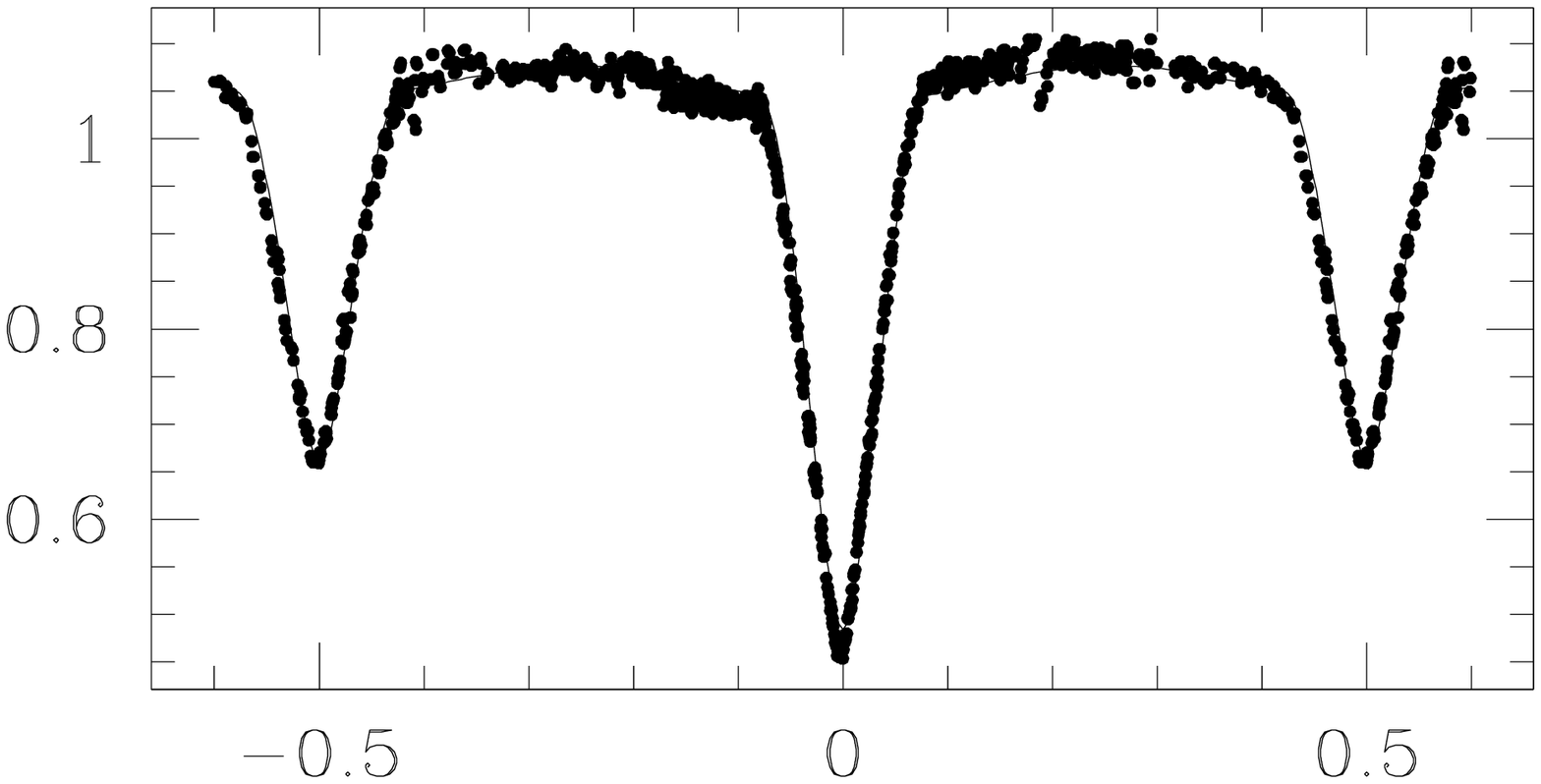}{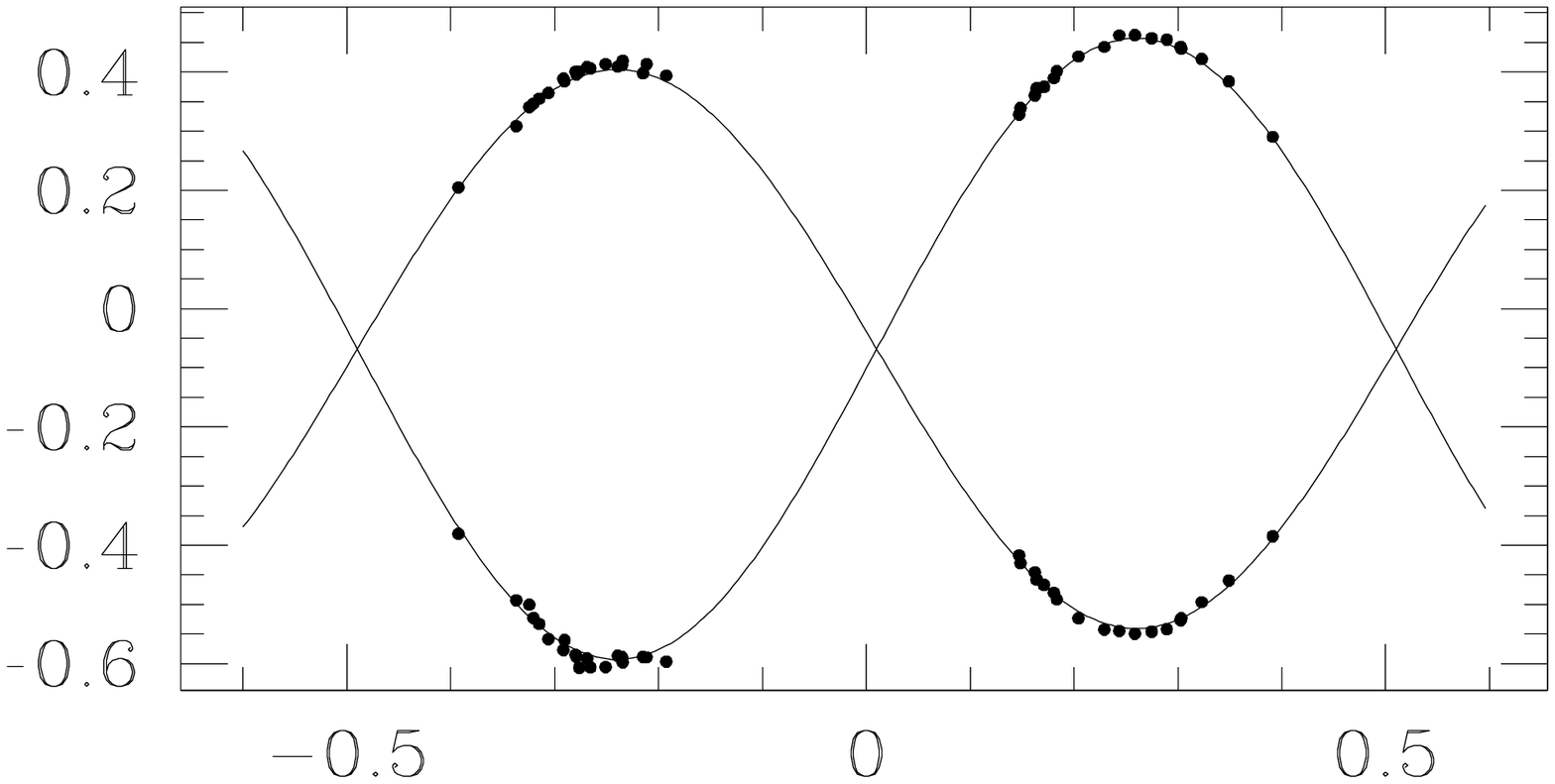} \\
\plottwo{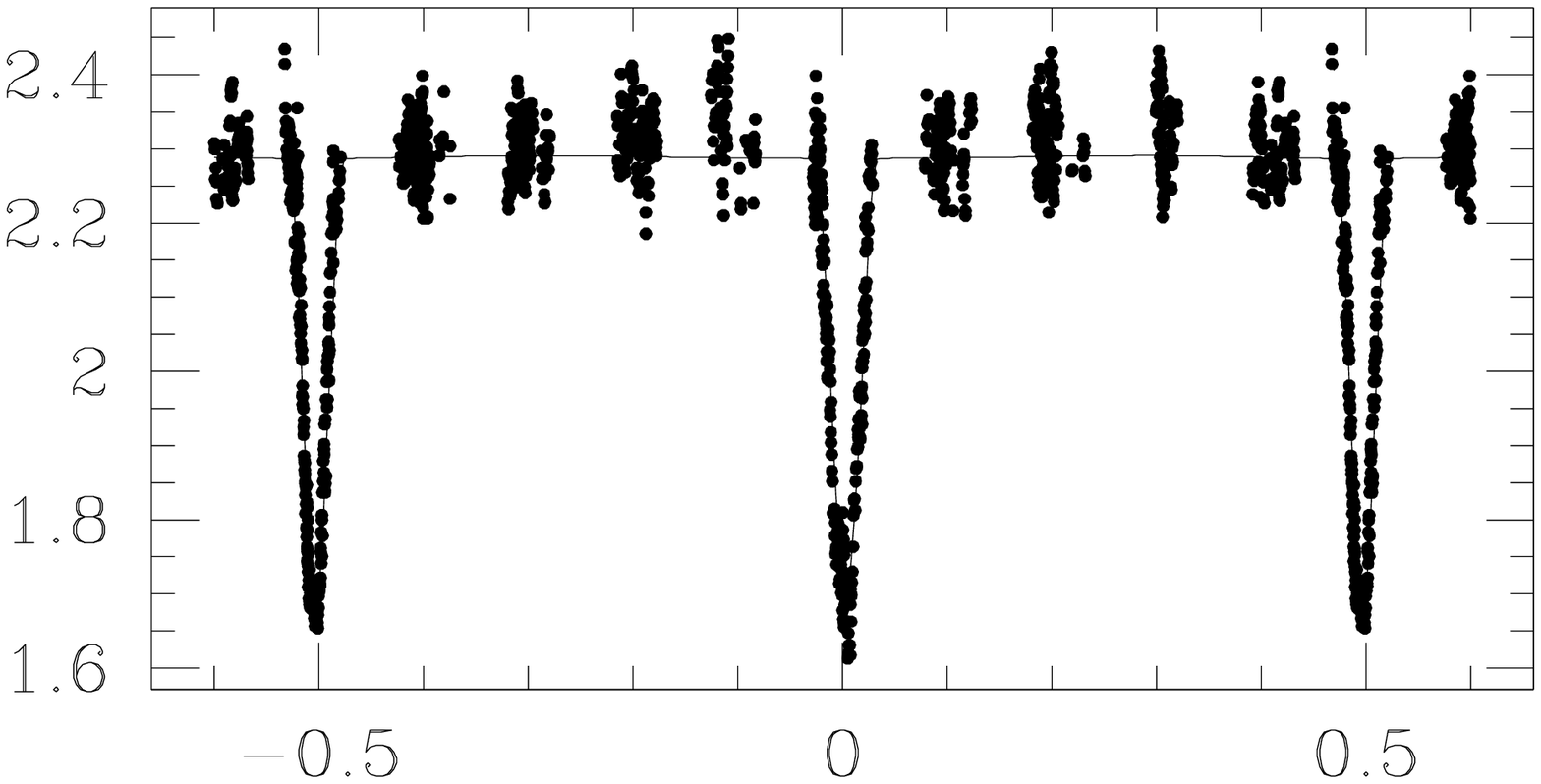}{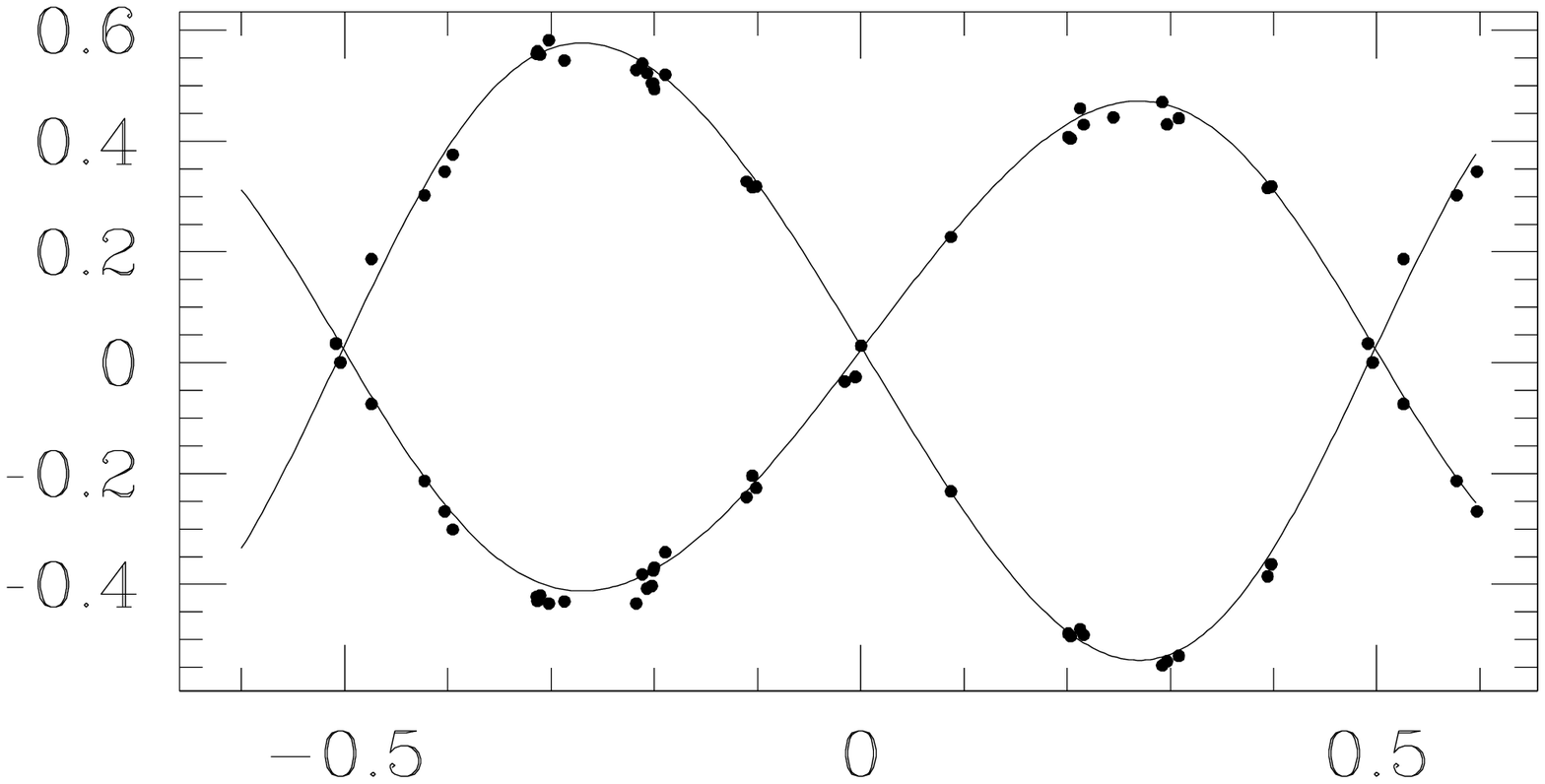} \\
\plottwo{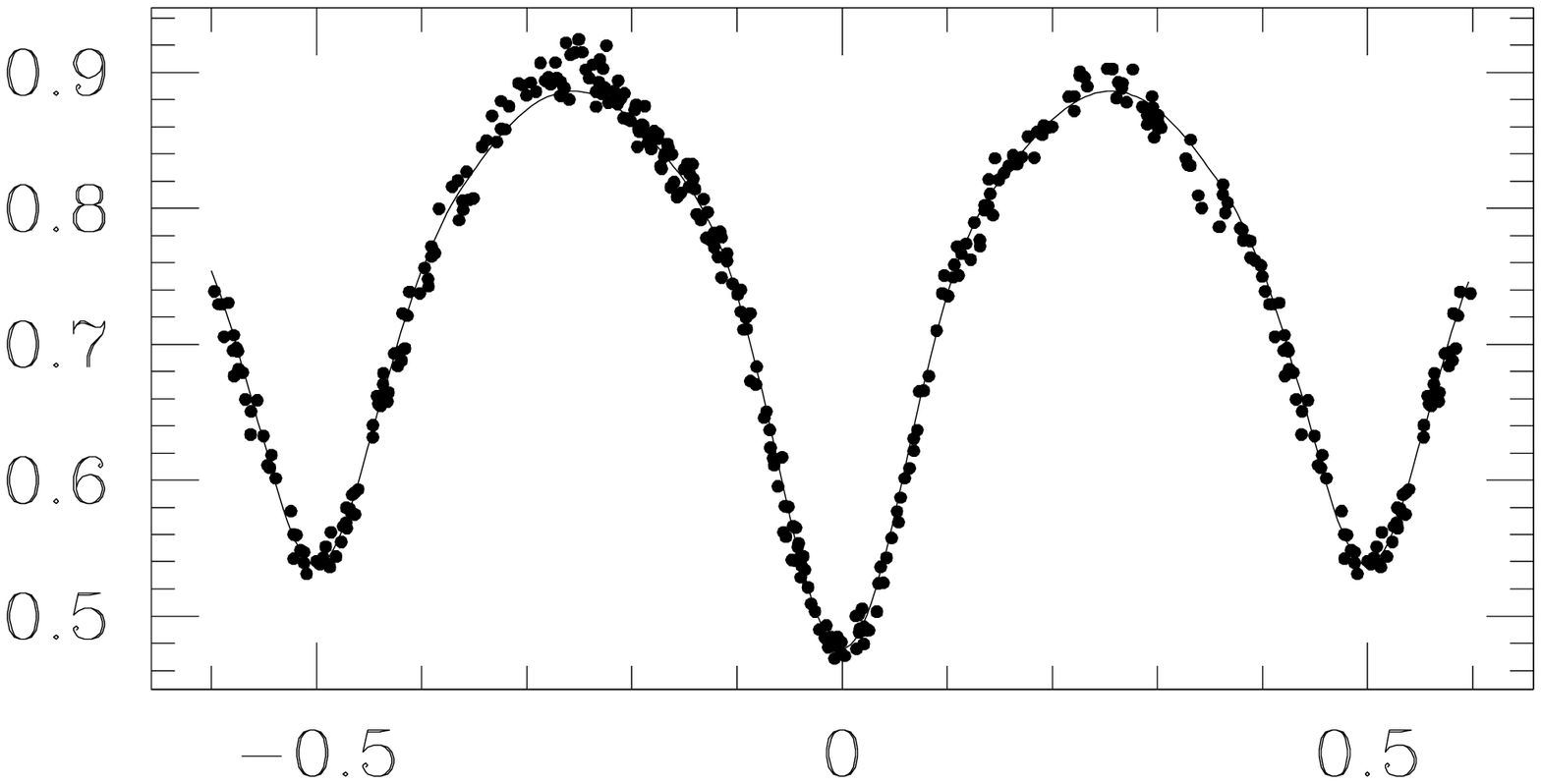}{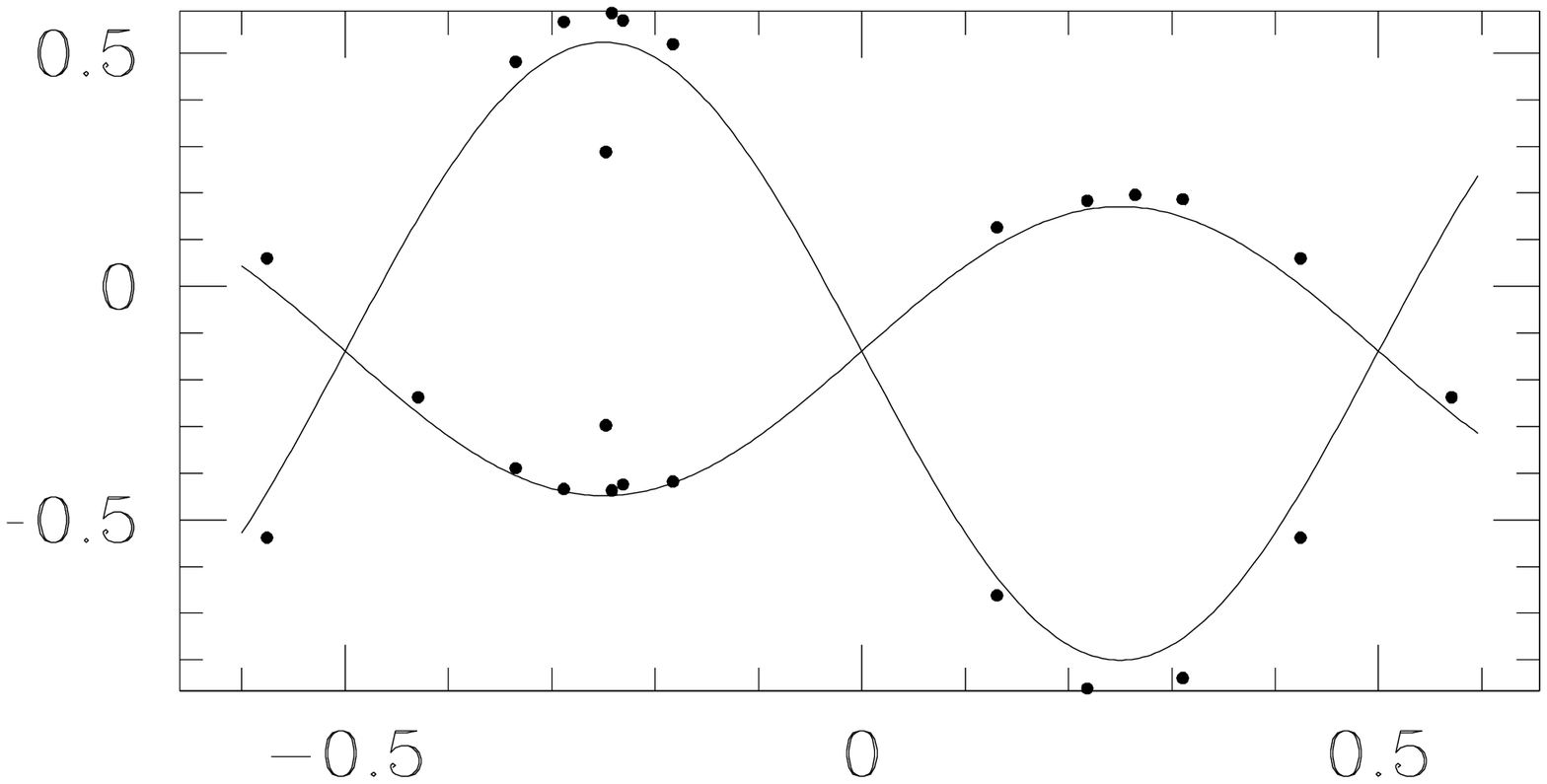} \\
\plottwo{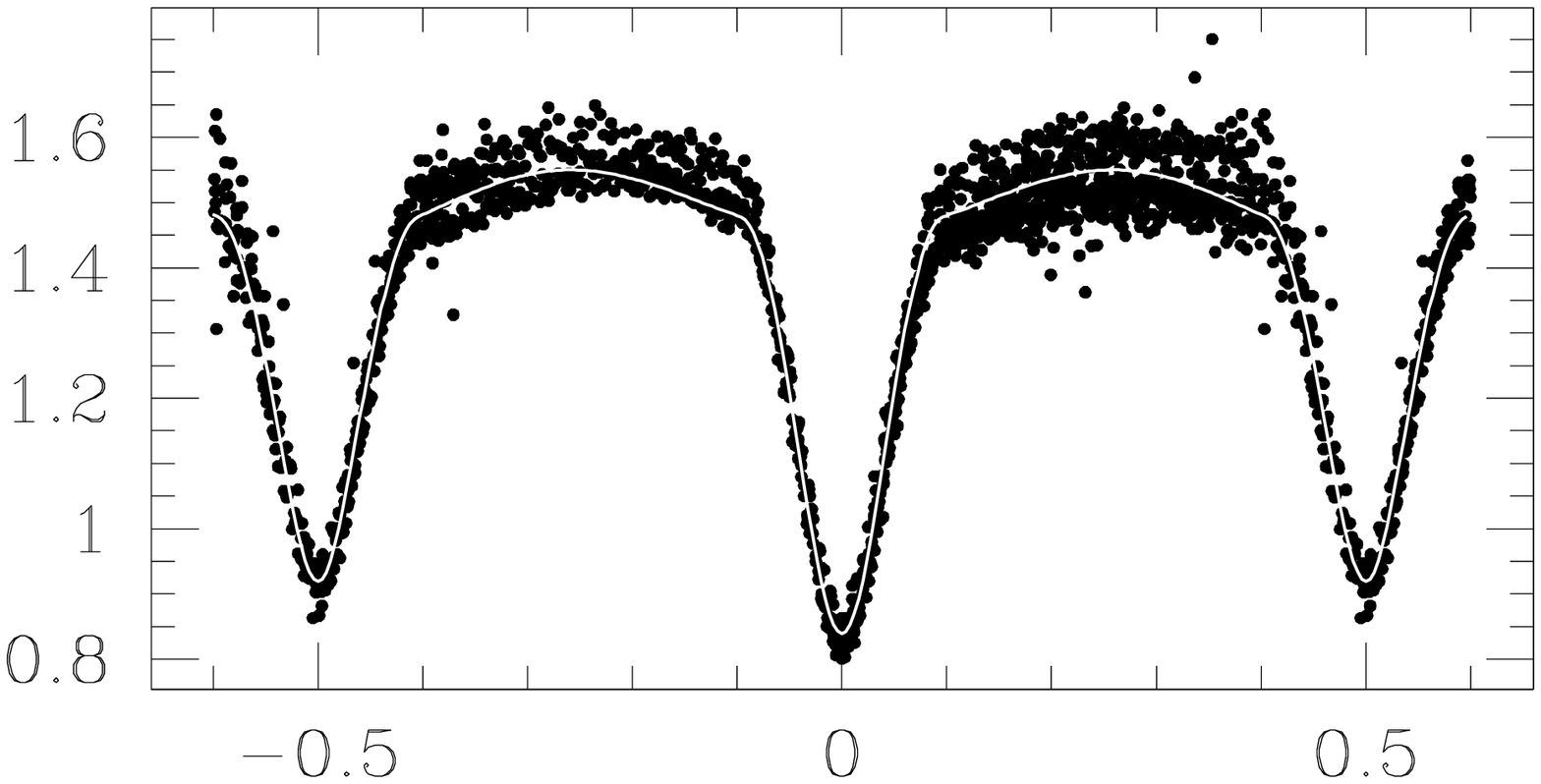}{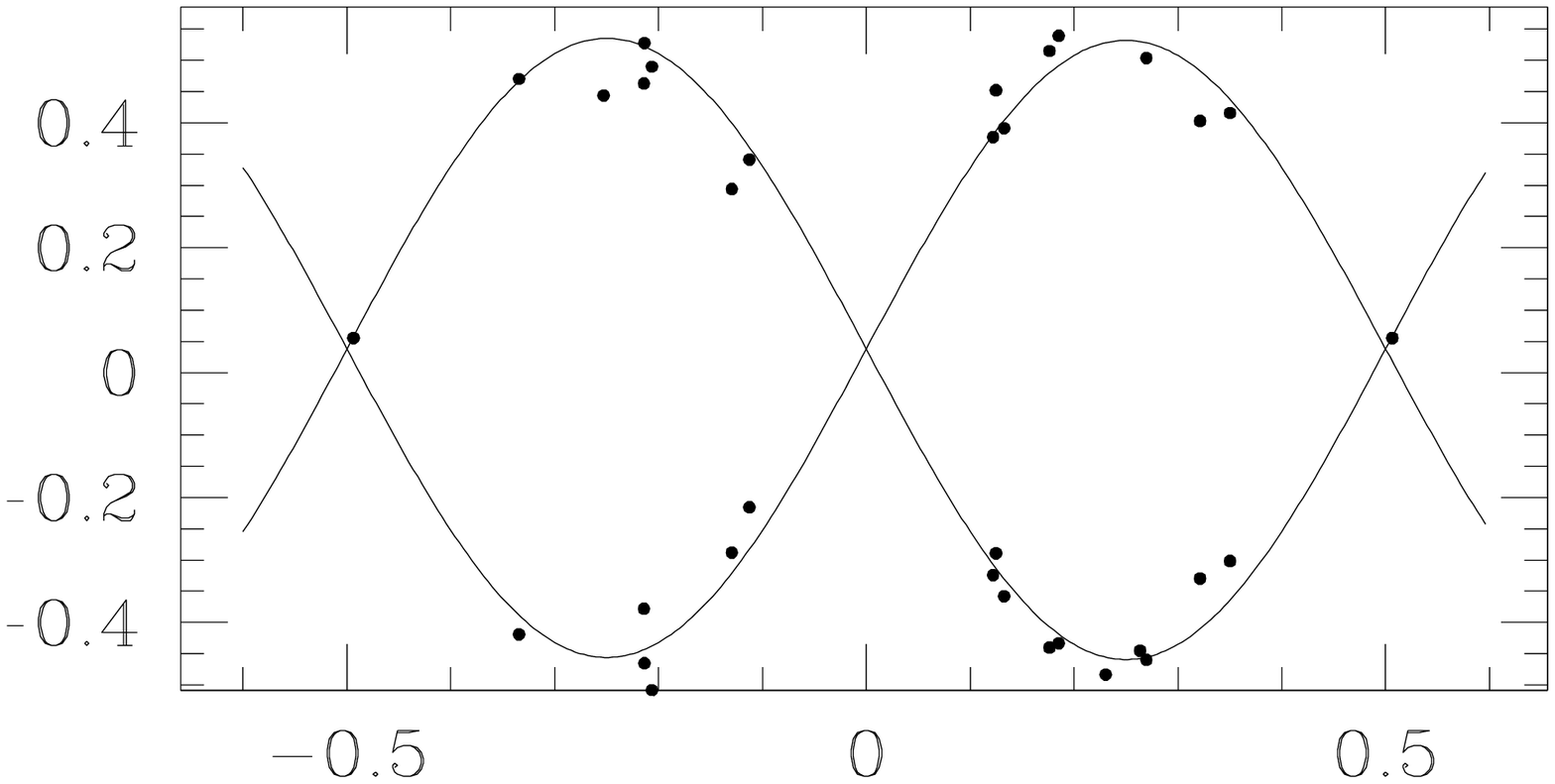} \\
\plottwo{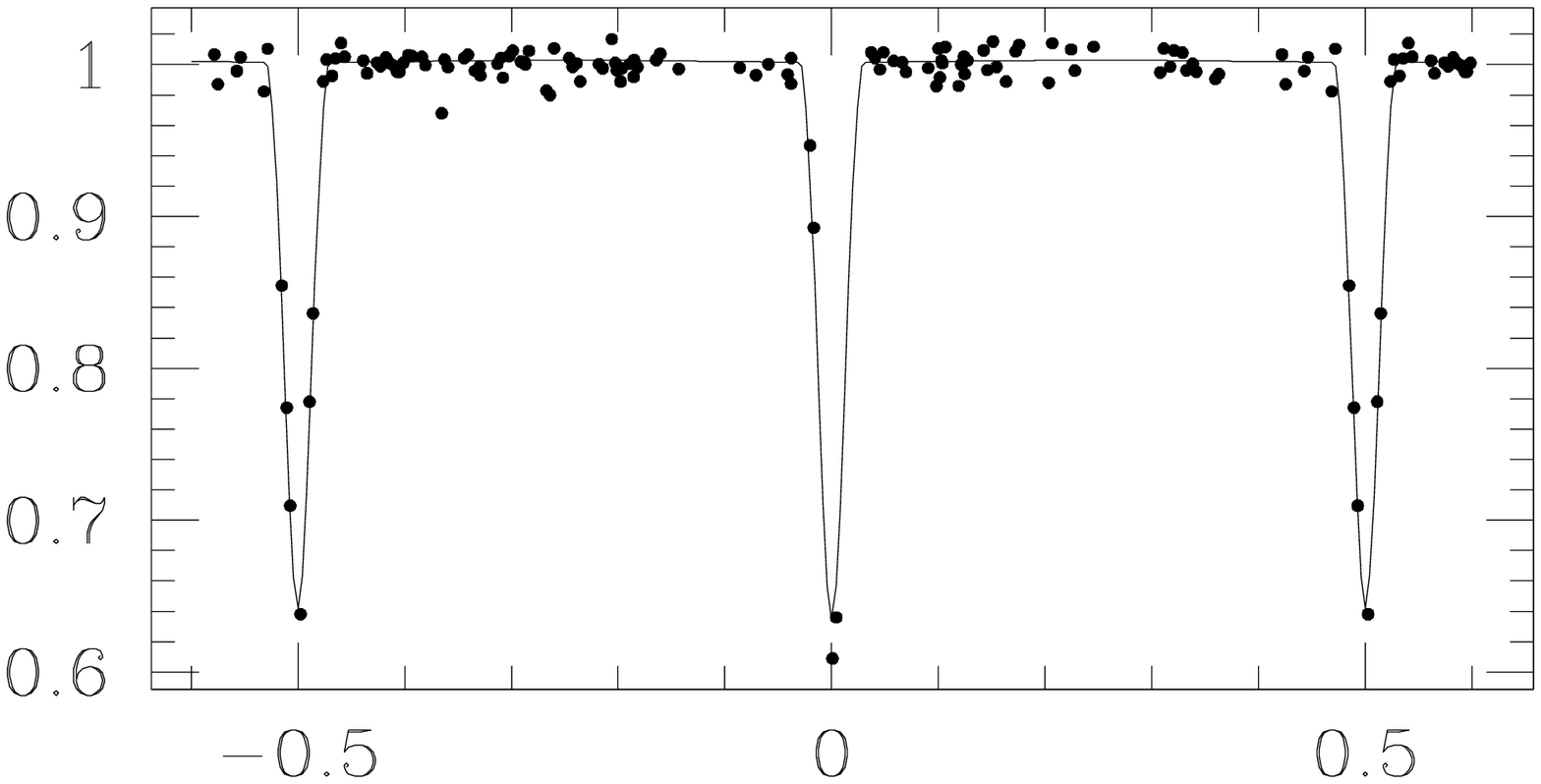}{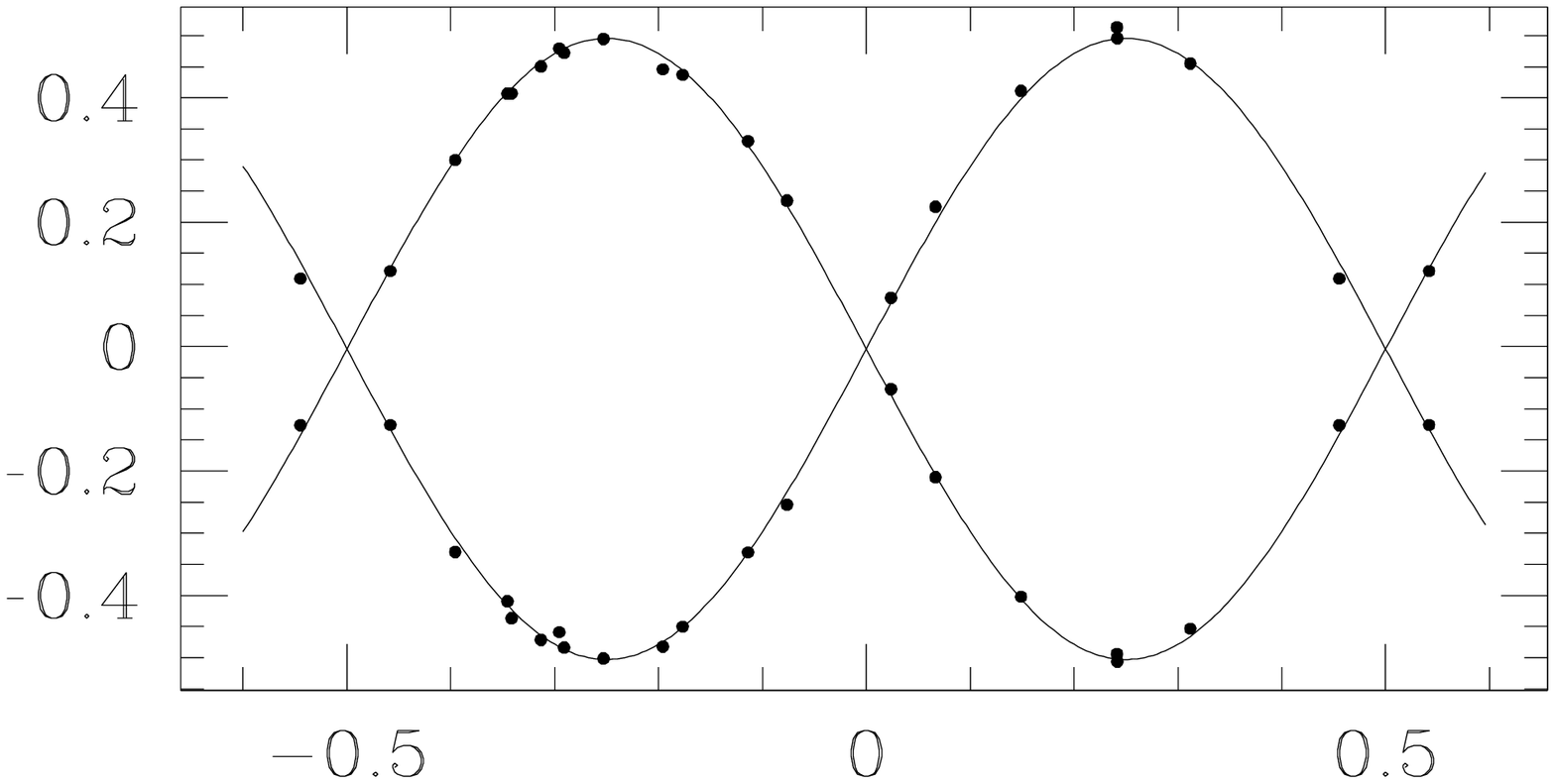} \\
\caption {Light curves (left) and radial velocity curves (right) of the 5 stars given in Table 1: BH Vir, GK Dra, TY Boo, UV Leo and V505 Per, respectively from top. The independent variable is in all cases phase and the dependent variable is relative flux for light curves and normalized radial velocities to $2\pi a/P$ for radial velocity curves. Measurements are marked with filled circles and the synthetic solution from Table 2 with a solid line.}
\end{figure}

\noindent
automatically retrieved solutions may be dangerous. Since we can recognize this problem, we may avoid it by using more adequate scripts, but there may be many caveats we must still anticipate.

\vspace{-3mm}
\begin{table}
\begin{center}
\small
\begin{tabular}{cccccccc}
\hline
\rule[-0.9ex]{0pt}{3ex}
Star Name & $a [R_\odot]$ & $q$ & $T_2/T_1$ & $i [{}^o]$ & $\Omega_1$ & $\Omega_2$ & $v_\gamma [km/s]$ \\
\hline
BH Vir    &  4.7(3) & 0.90(0) & 0.92(0) & 85.9(1) &  5.2(0) &  4.7(0) &  -0.2(8)  \\
GK Dra    & 28.9(4) & 1.26(0) & 1.01(0) & 83.9(2) & 10.5(1) & 10.5(1) &   1.7(12) \\
TY Boo    &  2.0(2) & 2.15(1) & 1.00(0) & 76.4(1) &  4.0(4) &  4.0(4) & -43.8(15) \\
UV Leo    &  4.0(3) & 1.00(1) & 0.97(1) & 84.1(1) &  5.1(0) &  4.4(0) &  12.8(6)  \\
V505 Per  & 15.1(1) & 0.98(1) & 0.99(0) & 87.0(2) & 11.5(2) & 11.2(2) &  -0.7(6) \\
\hline
\end{tabular}
\normalsize
\end{center}
\vspace{-3mm}
\caption{Computed values of basic stellar parameters using automative scripting. Standard deviation is given in parentheses as a measure of accuracy of the last decimal place. These are modeling standard deviations and may be larger in physical sense (Wilson 1998).}
\end{table}

\vspace{-7mm}
\begin{table}
\begin{center}
\small
\begin{tabular}{cccccccc}
\hline
\rule[-0.9ex]{0pt}{3ex}
Star Name & $a [R_\odot]$ & $q$ & $T_2/T_1$ & $i [{}^o]$ & $\Omega_1$ & $\Omega_2$ & $v_\gamma [km/s]$ \\
\hline
BH Vir    &  4.7 & 0.89 & 0.91 & 87.5 & 4.7  & 4.8  & 0.0   \\
GK Dra    & 29.0 & 1.26 & 0.98 & 85.8 & 14.1 & 12.2 & 3.8   \\
TY Boo    & 2.32 & 2.13 & 1.00 & 77.5 & 5.4  & 5.4  & -38.7 \\
UV Leo    & 3.91 & 0.96 & 0.98 & 82.6 & 4.6  & 4.3  & 13.0  \\
V505 Per  & 15.0 & 0.98 & 0.98 & 87.8 & 11.5 & 11.2 &  0.0  \\
\hline
\end{tabular}
\normalsize
\end{center}
\vspace{-3mm}
\caption{Adopted values of basic stellar parameters from references given in Table 1.}
\end{table}

The results demonstrated in the previous section are encouraging; however, one has to keep a fair amount of scientific scepticism about the adopted automation philosophy. This study was performed with an extremely undersampled statistics. A thorough study over large numbers of diverse eclipsing binaries is compulsory to gain confidence in the automation procedure. At this time we are still uncertain about the universality of the developed algorithm. We may be facing systematic errors due to conformity of our script: it \emph {may} invoke the existence of a preferred subspace within the parameter space, a dire consequence that would compromise all retrieved solutions.

\end{document}